\documentclass[aip,pra,twocolumn,showpacs]{revtex4}
\usepackage{graphicx,color}
\usepackage{amsmath}
\usepackage{longtable}
\begin{document}

\title{Fluid approach to evaluate sound velocity in Yukawa systems (complex plasmas)}

\author{Sergey A. Khrapak\footnote{Also at Joint Institute for High Temperatures, Russian Academy of Sciences, Moscow, Russia}
and Hubertus M. Thomas
 }
\date{\today}
\affiliation{Forschungsgruppe Komplexe Plasmen, Deutsches Zentrum f\"{u}r Luft- und Raumfahrt,
Oberpfaffenhofen, Germany}

\begin{abstract}
The conventional fluid description of multi-component plasma, supplemented by an appropriate equation of state for the macroparticle component, is used to evaluate the longitudinal sound velocity of Yukawa fluids. The obtained results are in very good agreement with those obtained earlier employing the quasi-localized charge approximation and molecular dynamics simulations in a rather broad parameter regime. Thus, a simple yet accurate tool to estimate the sound velocity across coupling regimes is proposed, which can be particularly helpful in estimating the dust-acoustic velocity in strongly coupled dusty (complex) plasmas. It is shown that, within the present approach, the sound velocity is completely determined by particle-particle correlations and the neutralizing medium (plasma), apart from providing screening of the Coulomb interaction, has no other effect on the sound propagation. The ratio of the actual sound velocity to its ``ideal gas'' (weak coupling) scale only weakly depends on the coupling strength in the fluid regime, but exhibits a pronounced decrease with the increase of the screening strength. The limitations of the present approach in applications to real complex plasmas are briefly discussed.
\end{abstract}

\maketitle

\section{Introduction}

Since the theoretical discussion of low-frequency dust-acoustic-waves (DAW)~\cite{Rao} and their experimental discovery~\cite{Chu,Barkan}, the concept of dust-acoustic velocity became one of the central concepts in the physics of dusty (complex) plasmas. The observations of DAWs and measurements of dust-acoustic (sound) velocities have been used to obtain useful information
about dusty plasmas systems under various conditions~\cite{Praburam,Thompson,Merlino1998,Fortov2000,Fortov2003,RatynskaiaIEEE,Fortov2004,Piel2006,Trottenberg,EThomas2007,Williams,EThomas2010} (we mainly concentrate on three-dimensional particle clouds here, which is also reflected in the reference list). In particular, experimentally measured DAW dispersion relations and sound velocities have been repeatedly used to estimate the electrical charge of particles in dusty plasmas under laboratory and microgravity conditions~\cite{KhrapakPoP2003,Yaroshenko2004,Ratynskaia2004,Khrapak2005,Schwabe2011}, employing various experimental techniques. For other aspects of DAWs and their significance for the field of dusty (complex) plasmas see, for instance, Refs.~\cite{ShuklaRMP,FortovBook,Merlino2014} and references therein.

Due to the large charge on the dust particles, complex plasmas often occur in a strongly coupled (liquid) state~\cite{FortovUFN,FortovPR,Bonitz}. The effects of strong coupling were neglected in the original derivation of DAW~\cite{Rao}, although the dispersion relation of the dust waves can clearly be seriously modified at strong coupling. Not surprisingly, the influence of strong coupling has been considered by many researchers, from the theoretical point of view, using different approaches. These include quasi-localized charge approximation (QLCA)~\cite{Rosenberg1997,KalmanPRL2000,Donko2008,Rosenberg2014}, generalized hydrodynamics~\cite{Kaw1998,Kaw2001}, local field correction description~\cite{MurilloPoP1998}, and ``multicomponent kinetic theory''~\cite{MurilloPoP2000}. Molecular dynamics (MD) simulations have also been performed to obtain wave dispersion relations in strongly coupled Yukawa fluids~\cite{MD1,MD2}. From experimental point of view, the topic has also received some attention~\cite{Merlino2014,Goree1996,Bandy}. Overall, the effect of strong coupling on weave phenomena in complex plasmas remains a very important current research topic~\cite{Merlino2014}.

The purpose of this paper is to investigate to which extent the standard fluid description of multi-component plasma, supplemented by an appropriate equation of state for the particle component, can be used for quantitative estimations of the longitudinal sound velocity in Yukawa fluids. It is known that the simplistic fluid treatment cannot reproduce the entire dispersion curve of longitudinal waves at strong coupling, and becomes particularly irrelevant in the regime of short wavelength. On the other hand, it can be expected to become more adequate for long wavelengths and low frequencies, i.e. in the regime of our present interest. There is also a direct evidence of the applicability of the conventional hydrodynamic description to sufficiently long waves in Yukawa systems, which comes from numerical simulations~\cite{Mithen2011,Com1}. In the simplest fluid formulation the effects of strong coupling are entirely accounted for by an appropriate equation of state. Here we make use of the recently proposed simple and accurate practical expressions for the internal energy and pressure of Yukawa fluids, applicable across coupling regimes~\cite{PExpr}. This allows us to obtain quantitative results and to perform detailed comparison with the results from other approaches.

In addition, we discuss another relevant issue, concerning the relation between the sound velocity in the conventional Yukawa system (charged particles interacting via Yukawa potential immersed into neutralizing medium) and that in a single component Yukawa system (an imaginary one-component system of particles interacting via the model Yukawa potential without any neutralizing medium). This question, from somewhat different perspective, has been recently considered in Ref.~\cite{Salin}.

The system considered in the following represents a collection of point-like highly charged particles in the neutralizing medium composed of plasma electrons and ions.
The particles interact via the pairwise Yukawa (or Debye-H\"{u}ckel) interaction potential
\begin{equation}\label{Yukawa}
V(r)= (Q^2/r)\exp(-r/\lambda_{\rm D}),
\end{equation}
where $Q$ is the particle charge, $\lambda_{\rm D}$ is the Debye screening length, and $r$ is the distance between a pair of particles. Screening is produced by the neutralizing plasma and the screening length, in the linear approximation, is given by $\lambda_{\rm D}=\lambda_{\rm Di}/\sqrt{1+(T_{\rm i}n_{\rm e}/T_{\rm e}n_{\rm i})}$, where $\lambda_{\rm Di}=\sqrt{T_{\rm i}/4\pi e^2 n_{\rm i}}$ is the ion Debye radius.
Here $n_{\rm i(e)}$ and $T_{\rm i(e)}$  are the ion (electron) density and temperature, respectively, and $e$ is the elementary charge. Since quasineutrality implies $n_{\rm i}\gtrsim n_{\rm e}$ (for negatively charged particles) and the electron temperature is much higher than that of the ions ($T_{\rm e}\gg T_{\rm i}$) in many practical situations, the screening is mostly associated with the ion component.

It is conventional to characterize the state of Yukawa systems in terms of two dimensionless parameters. The first is the coupling parameter, $\Gamma=Q^2/aT$, where $a= (3/4\pi n_{\rm{p}})^{1/3}$ is the Wigner-Seitz radius and $T$ is the kinetic temperature of the particle component (temperatures are measured in energy units throughout the paper). This is roughly the ratio of the (bar Coulomb) interaction energy between neighboring particles to their kinetic energy. The second is the screening parameter $\kappa=a/\lambda_{\rm D}$, which is roughly the ratio of the interparticle separation to the screening length.

Clearly, the idealized model described above oversimplifies considerably the actual rather complex interactions between the particles in real complex plasmas~\cite{FortovBook,KhrapakCPP,Morf2012}, although some experimentally observed trends can be reproduced by this simple consideration, at least qualitatively. However, more important in the present context is that simplifications involved make it possible to perform a direct comparison with the results obtained using other approaches (such as QLCA and MD). The limitations of our simplified model to describe real complex plasmas will be briefly discussed towards the end of the paper.

The paper is organized as follows. In Section~\ref{fluid} simplest fluid description of Yukawa model systems is formulated and an expression for the sound velocity is derived. Thermodynamic quantities of Yukawa fluids, necessary to evaluate the sound velocity, are provided in Section~\ref{thermo_sec}. The obtained results for the sound velocity are then analyzed and benchmarked against previously published results in Section~\ref{analysis}. This is followed by concluding remarks in Section~\ref{conclusion}.

\section{Simplest fluid description}\label{fluid}

We adopt the simplest fluid description of multi-component plasmas, similar to that used in the original derivation of DAW dispersion relation in Ref.~\cite{Rao}. In this formulation
electrons and ions provide equilibrium neutralizing medium and are described by
\begin{eqnarray}
-en_{\rm i}\nabla\phi = T_{\rm i}\nabla n_{\rm i}, \label{ions}
\\
en_{\rm e}\nabla\phi = T_{\rm e}\nabla n_{\rm e}, \label{electrons}
\end{eqnarray}
where $\phi$ is the electric potential. Equations (\ref{ions}) and (\ref{electrons}) result in equilibrium Boltzmann relations for the ion and electron densities in the wave potential.

The continuity and momentum equation for the particle component are
\begin{equation}\label{continuity}
\frac{\partial n_{\rm p}}{\partial t}+\nabla(n_{\rm p}{\bf v}_{\rm p})=0,
\end{equation}
\begin{equation}\label{momentum}
\frac{\partial {\bf v}_{\rm p}}{\partial t}+({\bf v_{\rm p}}\cdot \nabla){\bf v_{\rm p}}=-\frac{Q\nabla\phi}{m_{\rm p}}-\frac{\nabla P}{m_{\rm p}n_{\rm p}},
\end{equation}
where $n_{\rm p}$ and ${\bf v}_{\rm p}$ is the particle density and velocity, $m_{\rm p}$ is the particle mass, and $P$ is the pressure associated with the particle component.

In the limit of long-wavelength perturbations (acoustic regime) the system is quasineutral,
\begin{equation}\label{quasineutrality}
e n_{\rm i} -e n_{\rm e} +Qn_{\rm p}=0.
\end{equation}

The standard linearization procedure is then applied to the set of equations (\ref{ions})-(\ref{quasineutrality}). We assume $n_{\rm j}=n_{\rm j 0}+n_{\rm j 1}$ ($\rm{j=e,i,p}$), $\phi=\phi_{1}$, ${\bf v}_{\rm p}={\bf v}_{\rm p 1}$, and $P=P_0+P_1$, where the quantities with the subscript ``1'' correspond to small perturbations in the sound wave. All perturbations are proportional to $\exp(i{\bf kr}-i\omega t)$, where  $\omega$ is the wave frequency, and ${\bf k}$ is the wave vector. Since the sound wave is adiabatic~\cite{FluidMech}, the small change in pressure $P_1$ is related to the small change in particle density $n_{\rm p 1}$ by  $P_1=(\partial P/\partial n_{\rm p})_{S}n_{\rm p 1}$, where $S$ is the system entropy. The adiabatic compressibility can be expressed in terms of the isothermal compressibility using the thermodynamic relation  $(\partial P/\partial n_{\rm p})_{S}=\gamma (\partial P/\partial n_{\rm p})_{T}$, where $\gamma=C_{\rm P}/C_{\rm V}$ is the adiabatic index. This   results in the dispersion relation of the acoustic type
\begin{equation}\label{s1}
c_{\rm s}^2\equiv\frac{\omega^2}{k^2}=\omega_{\rm p}^2\lambda_{\rm D}^2+v_{\rm Tp}^2\gamma\mu,
\end{equation}
where $c_{\rm s}$ is the sound velocity, $\omega_{\rm p}=\sqrt{4\pi Q^2 n_{\rm p}/m_{\rm p}}$ is the plasma-particle frequency, $v_{\rm Tp}=\sqrt{T/m_{\rm p}}$ is the particle thermal velocity, and $\mu=(1/T)(\partial P/\partial n_{\rm p})_T$ is the isothermal compressibility modulus. In the limit of cold particle system ($T=0$) equation (\ref{s1}) reduces to the conventional dust-acoustic velocity
\begin{equation}\label{s0}
c_0=\omega_{\rm p}\lambda_{\rm D}=\sqrt{\frac{ZT_{\rm i}}{m_{\rm p}}}\sqrt{\frac{H\tau}{1+\tau+H\tau}},
\end{equation}
derived originally by Rao, Shukla and Yu~\cite{Rao} (see also ~\cite{FortovUFN}). Here $Z=|Q/e|$ is the particle charge number, $\tau=T_{\rm e}/T_{\rm i}$ is the electron-to-ion temperature ratio, and $H=Zn_{\rm p}/n_{\rm e}$ is the so-called Havnes parameter. Note that normally the sound velocity $c_{\rm s}$ is much larger than the particle thermal velocity $v_{\rm Tp}$ due to a large factor $Z$. Thus, the assumption of cold particles is justified at weak coupling, since in this regime $\mu \sim 1$ and $\gamma\sim \mathcal{O}(1)$. (Note, however, that strictly speaking the assumption of weak coupling and the condition of cold particles are not consistent, since $\Gamma\rightarrow\infty$ when $T\rightarrow 0$). The focus of the present study is mostly on the regime, when $\mu$ can considerably deviate from unity.

Rewritten in terms of reduced Yukawa state variables, $\kappa$ and $\Gamma$, the dispersion relation (\ref{s1}) becomes
\begin{equation}\label{s2}
c_{\rm s}=\omega_{\rm p}a\left(\frac{1}{\kappa^2}+\frac{\gamma\mu}{3\Gamma}\right)^{1/2}.
\end{equation}
This is identical to Eq. (27) from Ref.~\cite{DHH} in the limit of vanishing wavenumbers (long wavelengths).
The remaining step to identify the influence of strong coupling on the sound velocity is to take the appropriate values for $\gamma$ and $\mu$. The proper choice is discussed in the next Section.

\section{Thermodynamic functions of Yukawa fluids}\label{thermo_sec}

Main thermodynamic quantities of interest here are the internal energy $U$, Helmholtz free energy $F$, pressure $P$, specific heats $C_{\rm P}$ and $C_{\rm V}$, and the isothermal compressibility modulus $\mu=T^{-1}(\partial P/\partial n_{\rm p})_{T}$. If the internal energy is known, the following thermodynamic identities can be used to obtain other quantities~\cite{Landau}:
\begin{subequations}\label{thermo}
\begin{eqnarray}
U=-T^2\left(\frac{\partial}{\partial T}\frac{F}{T}\right)_V,
\\
P=-\left(\frac{\partial F}{\partial V}\right)_T,
\\
C_{\rm V}=\left(\frac{\partial U}{\partial T}\right)_V,
\\
C_{\rm P}-C_{\rm V}=-T\frac{(\partial P/\partial T)_V^2}{(\partial P/\partial V)_T}.
\end{eqnarray}
\end{subequations}
It is convenient to work with reduced units:  $u=U/NT$, $f=F/NT$, and $p=PV/NT$, $c_{\rm p}=C_{\rm P}/N$, $c_{\rm v}=C_{\rm V}/N$ and to express all the derivatives in terms of Yukawa system phase state variables, $\kappa$ and $\Gamma$. To do that we assume that the unperturbed electron and ion densities are not related to the particle density and their temperatures are fixed. This implies $\Gamma\propto (aT)^{-1}\propto T^{-1}n_{\rm p}^{1/3}$ and $\kappa\propto a\propto n_{\rm p}^{-1/3}$. This results in
\begin{displaymath}
\frac{\partial \Gamma}{\partial T}=-\frac{\Gamma}{T}, \quad \frac{\partial \Gamma}{\partial n_{\rm p}}=\frac{1}{3}\frac{\Gamma}{n_{\rm p}}, \quad \frac{\partial \kappa}{\partial T}= 0, \quad \frac{\partial \kappa}{\partial n_{\rm p}}=-\frac{1}{3}\frac{\kappa}{n_{\rm p}}.
\end{displaymath}

Thermodynamic identities (\ref{thermo}) yield the following relations between the reduced thermodynamic functions in $(\kappa,\Gamma)$ variables:
\begin{subequations}
\begin{eqnarray}
f(\kappa,\Gamma)=f_{\rm id}+\int_0^{\Gamma}d\Gamma'\left[u(\kappa, \Gamma')-3/2\right]/\Gamma',
\\
p(\kappa,\Gamma)=\frac{\Gamma}{3}\frac{\partial f}{\partial \Gamma}-\frac{\kappa}{3}\frac{\partial f}{\partial \kappa},
\end{eqnarray}
\end{subequations}
where $f_{\rm id}=\ln\left[(2\pi\hbar^2/m_{\rm p}T)^{3/2}n_{\rm p}\right]-1$ is the ideal gas contribution to the free energy.
The specific heat per particle at constant volume is expressed in terms of the reduced internal energy,
\begin{equation}
c_{\rm v} (\kappa,\Gamma)=u-\Gamma (\partial u/\partial \Gamma).
\end{equation}
The isothermal compressibility modulus is related to the reduced pressure via
\begin{equation}\label{compr}
\mu(\kappa,\Gamma)= p+\frac{\Gamma}{3}\frac{\partial p}{\partial \Gamma}-\frac{\kappa}{3}\frac{\partial p}{\partial \kappa}.
\end{equation}
Finally, the difference between reduced specific heats at constant pressure and volume is
\begin{equation}
c_{\rm p}-c_{\rm v} = \frac{[p-\Gamma(\partial p/\partial \Gamma)]^2}{\mu}.
\end{equation}
Note that it is conventional to decompose the thermodynamic quantities into the ideal gas contribution and excess contribution, associated with particle-particle correlations. We have not done this here in order to simplify the notation. The ideal gas contributions to the reduced energy and pressure are $u_{\rm id}=3/2$ and $p_{\rm id}=1$, respectively.

The multi-component character of the system under consideration implies that the thermodynamic quantities considered above can contain two distinct contributions. The first comes from particle-particle interactions and characterizes a single component Yukawa system without neutralizing background (it also contains the ideal gas contribution for non-interacting particles). The second is the plasma-related contribution, describing plasma and plasma-particle interactions. In particular, the plasma-related contribution to the internal energy of the system is~\cite{Ham94}
\begin{equation}
u_{\rm pl}=-\frac{3\Gamma}{2\kappa^2}-\frac{\kappa\Gamma}{2}.
\end{equation}
The first term represents the (free) energy of the electron-ion plasma that, on average, neutralizes the
charge of the particles, while the second term gives the (free) energy of the plasma sheath around each particle, in the linear approximation. This latter term does not affect neither pressure, nor the compressibility modulus (and hence there is no effect on specific heats)~\cite{DHH,ISM}. It results, however, in the polarization force acting on the particles in non-uniform plasmas, see Appendix~\ref{Individual}. The plasma-related contributions to $p$ and $\mu$ are thus~\cite{PExpr}
\begin{equation}
p_{\rm pl}=-3\Gamma/2\kappa^2 \quad {\rm end}\quad \mu_{\rm pl}= -3\Gamma/\kappa^2.
\end{equation}
The necessity to account for the plasma-related contribution can be easily understood realizing that it ensures the expected ideal gas values $p=1$ and $\mu=1$ in the high-temperature limit, when particle-particle correlations are completely absent.

Thermodynamic properties of Yukawa systems have been extensively studied using various computational and analytical techniques. Relevant examples include Monte Carlo (MC) and molecular dynamics (MD) numerical simulations~\cite{Robbins,Meijer,Hamaguchi,Caillol}, as well as integral equation theoretical studies~\cite{Tejero,Kalman2000,Faussurier,SMSA}. Accurate values of various thermodynamic quantities have been tabulated in a wide region of $(\kappa,\Gamma)$ phase space (in particular, see Refs.~\cite{Hamaguchi,Caillol}). Simple practical expressions for $u$, $p$, and $\mu$ have been put forward recently~\cite{PExpr}. These expressions stem from the original observation by Rosenfeld and Tarazona~\cite{Rosenfeld1998,Rosenfeld2000} that the thermal component of the internal energy of various soft repulsive systems (including the Yukawa case) exhibits quasi-universal dependence on the properly normalized coupling parameter. The resulting expressions are applicable in a wide range of coupling and demonstrate remarkable agreement with the results from numerical simulations. The expression for the energy of the single component Yukawa fluid suggested in~\cite{PExpr} is
\begin{equation}\label{uexpr}
u_{\rm pp}(\kappa,\Gamma)=\frac{3}{2}+ \epsilon + \frac{\kappa(\kappa+1)\Gamma}{(\kappa+1)+(\kappa-1)e^{2\kappa}}+\delta (\Gamma/\Gamma_{\rm m})^{2/5},
\end{equation}
where $\Gamma_{\rm m}$ denotes the coupling parameter at the fluid-solid phase transition and the subscript ``pp'' means that only the contribution coming from particle-particle correlations (which also includes the ideal gas term) is considered. The functional dependence $\Gamma_{\rm m}(\kappa)$ can be approximated by~\cite{VaulinaJETP,VaulinaPRE}
\begin{equation}\label{melt}
\Gamma_{\rm m}(\kappa)\simeq \frac{172 \exp(\alpha\kappa)}{1+\alpha\kappa+\tfrac{1}{2}\alpha^2\kappa^2},
\end{equation}
where the constant $\alpha=(4\pi/3)^{1/3}\simeq 1.612$  is the ratio of the mean interparticle distance $\Delta=n^{-1/3}$ to the Wigner-Seitz radius $a$. It has been also suggested~\cite{PExpr} to use $\delta=3.2$ and $\epsilon=-0.1$ in Eq.~(\ref{uexpr}).

\begin{figure}
\includegraphics[width=7.5cm]{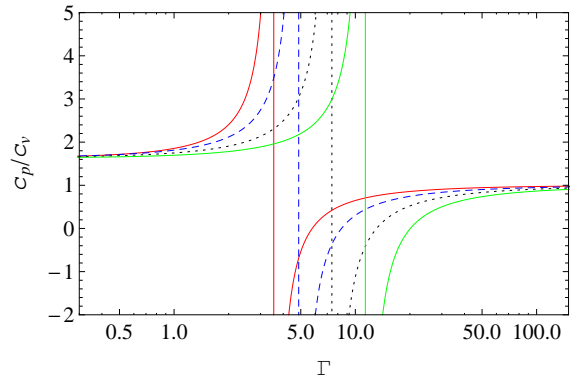}
\caption{(Color online) Adiabatic index $\gamma=c_{\rm p}/c_{\rm v}$ as a function of the coupling parameter $\Gamma$ for four values of the screening parameter: $\kappa = 1.0$ (red solid curve), $\kappa=2.0$ (blue dashed curve), $\kappa=3.0$ (black dotted curve), and $\kappa=4.0$ (green solid curve). The discontinuity in $\gamma$ is the consequence of plasma-related contribution to the thermodynamic quantities.}
\label{fig1}
\end{figure}

The expressions for $p$, $\mu$, and $\gamma$ can be easily derived using Eqs.~(\ref{uexpr})-(\ref{melt}), as has been done in Ref.~\cite{PExpr}. Note that when plasma-related contribution is properly accounted for, $p$ and $\mu$ tend to unity in the ideal gas limit (vanishing correlations) and become negative at sufficiently strong coupling. The adiabatic index $\gamma$ tends to  $5/3$ in the ideal gas limit, exhibit a discontinuity at moderate coupling (when $\mu=0$) and then rapidly approach the asymptote $\gamma=1$ at strong coupling. This is illustrated in Figure~\ref{fig1}, where the adiabatic index $\gamma$ is plotted as a function of the coupling parameter $\Gamma$ for four representative values of $\kappa$. The discontinuity in $\gamma$ may deserve special attention, but has no impact on the problem under consideration, because the product $\gamma\mu$ remains finite. It decreases monotonously from the ideal gas value $5/3$ to highly negative values when $\Gamma$ increases.
In this regime the sound velocity should decrease compared to its ideal gas value (evaluated at the same $\kappa$ and $\Gamma$ values), as becomes immediately obvious from Eq.~(\ref{s2}).

\section{Sound velocity}\label{analysis}

Now we can evaluate the sound velocity of Yukawa fluids in a broad range of coupling. First, we write
\begin{equation}\label{gm}
\gamma\mu = \mu+\frac{[p-\Gamma(\partial p/\partial \Gamma)]^2}{u-\Gamma (\partial u/\partial \Gamma)},
\end{equation}
where $u$, $p$, and $\mu$ account for both particle-particle correlation and plasma-related effects, i.e. $u=u_{\rm pp}+u_{\rm pl}$, $p=p_{\rm pp}+p_{\rm pl}$, and $\mu=\mu_{\rm pp}+\mu_{\rm pl}$.    The first important observation is that when substituting Eq.~(\ref{gm}) into the expression for the sound velocity (\ref{s2}), the plasma-related contribution to the isothermal compressibility modulus $\mu_{\rm pl}=-3\Gamma/\kappa^2$ {\it cancels out exactly} the plasma contribution to the dispersion relation given by the term $1/\kappa^2$ in (\ref{s2}). The expression for the sound velocity can be rewritten as
\begin{equation}\label{s3}
c_{\rm s}=\omega_{\rm p}a\left(\frac{\mu_{\rm pp}}{3\Gamma}+\frac{[p-\Gamma(\partial p/\partial \Gamma)]^2}{3\Gamma \left[u-\Gamma (\partial u/\partial \Gamma)\right]}\right)^{1/2}.
\end{equation}
The next important observation is that since the plasma-related contributions $u_{\rm pl}$ and $p_{\rm pl}$ are both linear in $\Gamma$, they have no effect on the second term in brackets of Eq.~(\ref{s3}). Thus, the sound velocity of a system of charged particles immersed in the neutralizing plasma environment is equal to that of an imaginary single component Yukawa system. In other words, the magnitude of the sound velocity is completely determined by particle-particle correlations and the neutralizing medium only affects the interparticle interactions, but has no other effect on the sound propagation. Therefore, the expressions for $u_{\rm pp}(\kappa,\Gamma)$, $p_{\rm pp}(\kappa,\Gamma)$, and $\mu_{\rm pp}(\kappa,\Gamma)$ proposed in Ref.~\cite{PExpr} can be directly substituted into Eq.~(\ref{s3}) to evaluate the sound velocity for a given pair of $\kappa$ and $\Gamma$. We have done this, the obtained results are discussed below.

\begin{figure}
\includegraphics[width=7.5cm]{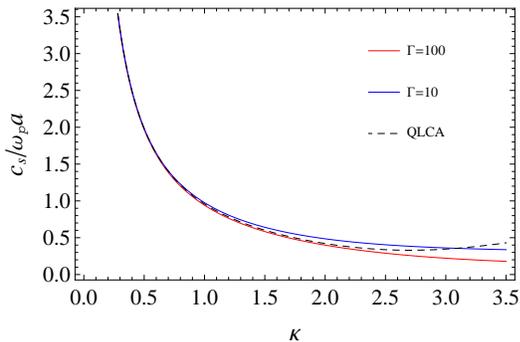}
\caption{(Color online) Reduced sound velocity of Yukawa fluids, $c_{\rm s}/\omega_{\rm p}a$, as a function of the screening parameter $\kappa$. The solid curves correspond to the results of the simple fluid approach of this paper for $\Gamma=10$ (blue curve) and $\Gamma=100$ (red curve). The dashed curve is plotted using QLCA result of Ref.~\cite{KalmanPRL2000}, given by Eqs.~(\ref{Kal1}) and (\ref{Kal2}).}
\label{fig2}
\end{figure}

First, it is convenient to benchmark our results against quantitative data published in previous works. For example, Kalman {\it et al}.~\cite{KalmanPRL2000} suggested the following expression for the longitudinal sound velocity, based on the results of QLCA model,
\begin{equation}\label{Kal1}
c_{\rm s}=\omega_{\rm p}a\left[1/\kappa^2+f(\kappa)\right]^{1/2},
\end{equation}
where
\begin{equation}\label{Kal2}
f(\kappa)\simeq -0.0799-0.0046\kappa^2+0.0016\kappa^4
\end{equation}
is a fitting function applicable for $\kappa<2.5$ (note, that in this approach $c_{\rm s}/\omega_{\rm p}a$ is a function of the screening parameter $\kappa$ alone). In Figure~\ref{fig2} we plot
the corresponding curve along with our calculation for the two values of the coupling parameter, $\Gamma=10$ and $\Gamma= 100$. The curve corresponding to stronger coupling is practically indistinguishable from the QLCA result of Ref.~\cite{Kalman2000}.

A more detailed comparison with the QLCA is provided in Table~\ref{Tab0}. The first two columns specify the location of the system in terms of screening parameter $\kappa$ and reduced coupling parameter $\Gamma/\Gamma_{\rm m}$. The third column lists the values of the reduced sound velocity obtained using the QLCA model in Ref.~\cite{Donko2008}. The last column contains the values of the reduced sound velocity calculated from the simple fluid  approach of this paper using Eq.~(\ref{s3}). Some weak dependence on the coupling strength is now present. The sound velocity obtained using the simple fluid approach is slightly smaller than that from the QLCA method, but the overall agreement is quite good. It is unlikely that the difference arises due to an approximate character of the equation of state employed here. The latter agrees with the accurate numerical data to within a tiny fraction of a percent in the regime $\kappa\lesssim 3$ and $\Gamma/\Gamma_{\rm m}\gtrsim 0.1$~\cite{PExpr}. Note that the QLCA model~\cite{Golden2000} is by construction more appropriate to describe high-frequency short-wavelength phenomena and as such it is not necessarily more accurate than the fluid approach in predicting the sound velocity.

\begin{table}
\caption{\label{Tab0} Reduced sound velocity $c_{\rm s}/\omega_{\rm p}a$ of Yukawa fluids as calculated from the QLC approximation and present fluid model for several phase state points. QLCA data are from Ref.~\cite{Donko2008}. For details see the text.}
\begin{ruledtabular}
\begin{tabular}{llcc}
$\kappa$ &  $\Gamma/\Gamma_{\rm m}$ &  QLCA & Fluid  \\ \hline
1.0 & 0.12 & 0.96 & 0.95 \\
1.0 & 0.70 & 0.96 & 0.94 \\
2.0 & 0.12 & 0.42 & 0.41 \\
2.0 & 0.70 & 0.41 & 0.39 \\
3.0 & 0.12 & 0.23 & 0.21 \\
3.0 & 0.70 & 0.21 & 0.19 \\

\end{tabular}
\end{ruledtabular}
\end{table}

\begin{figure}
\includegraphics[width=7.5cm]{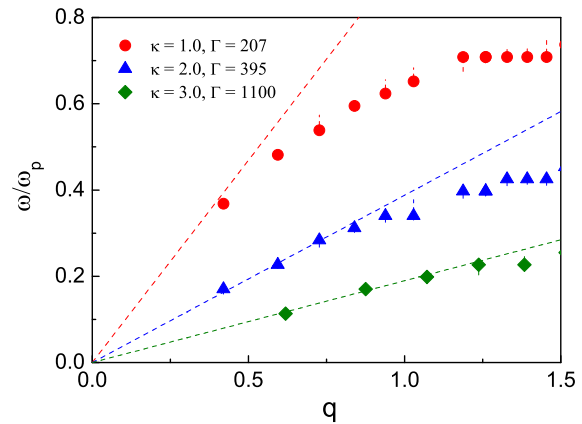}
\caption{(Color online) Long-wavelength dispersion of the longitudinal waves in Yukawa fluids near freezing, plotted in the $(q, \omega/\omega_{\rm p})$ plane, where $q=ka$ is the reduced wavenumber. Symbols correspond to the results from numerical simulations of Ref.~\cite{MD1} for $\kappa=1.0$ (red circles), $\kappa=2.0$ (blue triangles), and $\kappa=3.0$ (olive rhombuses). The corresponding dashed lines correspond to the acoustic asymptotes $\omega=kc_{\rm s}$ with the sound velocity $c_{\rm s}$ calculated using the fluid approach described in the present paper [Eq.~(\ref{s3})]. }
\label{fig3}
\end{figure}

In Figure~\ref{fig3} we compare the long-wavelength part of the longitudinal dispersion relation of Yukawa fluids obtained in a numerical (MD) experiment in Ref.~\cite{MD1} with the acoustic asymptote $\omega=kc_{\rm s}$, where $c_{\rm s}$ is calculated from the present fluid approach. The behavior of the numerically obtained dispersion curves is clearly consistent with the respective acoustic asymptotes (dashed lines). Moreover, Figure~\ref{fig3} demonstrates that at sufficiently strong screening (say $\kappa\gtrsim 2$), the acoustic asymptote describes well the dispersion curve up to $q\simeq 1$, where $q=ka$ is the reduced wavenumber. This would be sufficient for many experimental investigations of DAWs in complex plasmas.

\begin{figure}
\includegraphics[width=7.5cm]{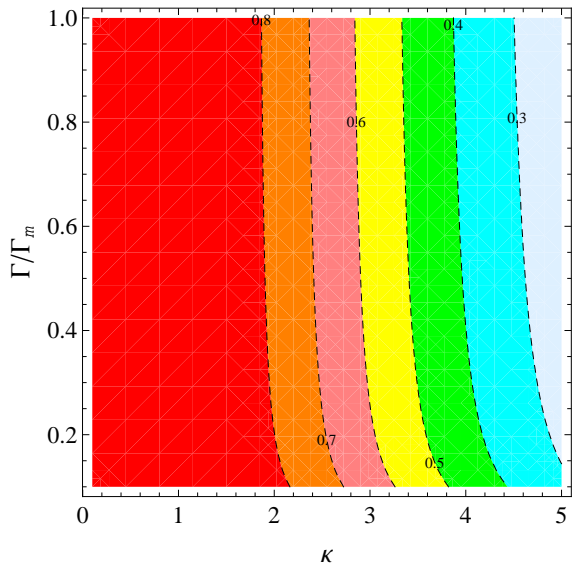}
\caption{(Color online) Contour plot of the reduced sound velocity $c_{\rm s}/c_0$ of Yukawa fluids in the plane $(\kappa, \Gamma/\Gamma_{\rm m})$. Calculations are made using the fluid model described in this paper.}
\label{fig4}
\end{figure}

Having benchmarked the present fluid results against previous results from the QLCA model and MD simulations, let us investigate the dependence of the sound velocity on coupling and screening in detail. It is particularly useful to analyze the behavior of the quantity $c_{\rm s}/c_{0}$, which is the ratio of the actual sound velocity of a Yukawa fluid to the respective limiting ``ideal gas'' (weak coupling limit) value given by Eq.~(\ref{s0}). The contour plot of this quantity in the $(\kappa, \Gamma/\Gamma_{\rm m})$ plane is shown in Fig.~\ref{fig4}. First, we observe only weak dependence of the quantity $c_{\rm s}/c_{0}$ on $\Gamma$ deep in the fluid regime (on approaching the fluid-solid phase transition). This implies that the absolute value of $c_{\rm s}$ increases with $\Gamma$, because $c_0\propto \sqrt{\Gamma}$. Second, we observe that the ratio $c_{\rm s}/c_{0}$ is sensitive to the screening parameter, and decreases as $\kappa$ increases. It drops by almost one order of magnitude on the way from the weakly screened regime $\kappa\lesssim 1$ to the strongly screened regime with $\kappa\simeq 5$.

\begin{figure}
\includegraphics[width=7.5cm]{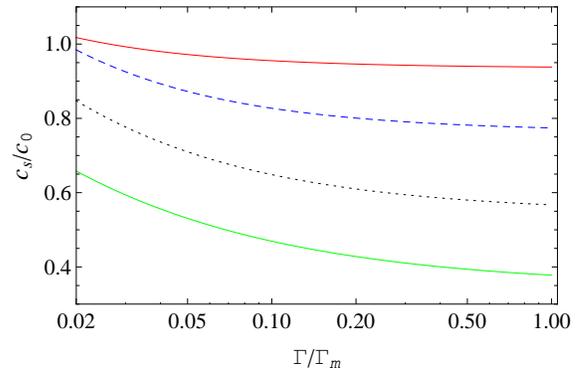}
\caption{(Color online) The reduced sound velocity $c_{\rm s}/c_0$ of Yukawa fluids versus the reduced coupling parameter $\Gamma/\Gamma_{\rm m}$. The curves (from top to bottom) correspond to $\kappa=1.0$, $\kappa=2.0$, $\kappa=3.0$, and $\kappa=4.0$, respectively. Calculations are made using the fluid model described in this paper. The equation of state used in these calculations may not be very accurate at $\Gamma/\Gamma_{\rm m}\lesssim 0.1$.}
\label{fig5}
\end{figure}

Figure~\ref{fig5} shows the dependence of $c_{\rm s}/c_0$ on the reduced coupling parameter for four values of the screening parameter (from $\kappa=1.0$ to $\kappa=4.0$). It demonstrates that the main drop in the reduced sound velocity occurs already in the regime of relatively weak coupling. In the strong coupling regime (say $\Gamma/\Gamma_{\rm m}\gtrsim 0.1$) the decrease is very slow, which in some sense justifies neglecting the $\Gamma$-dependence, like in Eq.~(\ref{Kal1}) above.

\section{Concluding remarks}\label{conclusion}

The main results obtained in this study can be summarized as follows. We have applied the standard fluid description of multi-component plasmas, supplemented by an appropriate equation of state,
to evaluate the sound velocity in Yukawa fluids. It turns out that the sound velocity is completely determined by particle-particle correlations. The obtained values of sound velocities are in rather good agreement with the previously published results obtained using QLCA approach and MD simulations. The main trends observed include slow decrease of the sound velocity of a Yukawa fluid compared to its ``ideal gas'' scale with increase in the coupling strength, and more pronounced decrease with increase of the screening strength. Overall, the standard fluid description, with a proper model for an equation of state, provides (perhaps not very surprisingly) simple
yet accurate tool to evaluate sound velocity in Yukawa fluids and related systems in a rather broad parameter regime.

Among the possible applications of the present results, the topics related to low-frequency wave phenomena in complex (dusty) plasmas seem particularly relevant. However, it is important to take into account the following circumstances, which can limit the applicability of the simple model discussed here in some practical situations. First of all, when deriving the dispersion relation in Section~\ref{fluid}, the most simple possible formulation of the problem has been employed. In real situations one may need to account for the presence of external electric fields and resulting drifts between different charged species, various kinds of collisions present in the system, effects of plasma production and loss, particle charge variations in the presence of the waves, additional forces acting on the particle component, etc. Most of these effects can be relatively easily included into the conventional fluid formalism, for some relevant examples we refer to Refs.~\cite{AngeloPoP98,IvlevPoP99,IvlevPoP2000,KhrapakPoP2003_1,KhrapakPRL2009,KhrapakPoP2010,YaroshenkoPoP2012,RuhunusiriPoP2014}. Thus, careful analysis of the most important processes in each concrete practical situation is required. Inclusion of these processes into consideration should not become a major problem from the theoretical point of view.

The second class of problems is related to the openness of the complex plasma systems. Plasma electrons and ions are continuously lost on the particle surface and the particle charge is set by the condition of no net electrical current to the surface (or, equivalently, floating potential at the particle surface). This is known to result in some deviations from the Yukawa-type potential around the particles~\cite{FortovBook,FortovPR} and, therefore, some deviations from the thermodynamic functions of conventional Yukawa fluids can also be expected. Perhaps even more important is that the particle charge in complex plasmas is not fixed, but depends on the parameters of the surrounding plasma. In particular, the charge becomes a function of the particle density via the so called ``charge cannibalism'' effect~\cite{Havnes1984,BarkanPRL1994,KhrapakEPL2010}. This effect operates as follows: When the particle density increases, the negative charge carried by the particle component also increases, which results in some reduction of the electron-to-ion density ratio (electron depletion) in view of the quasineutrality condition. In turn, this suppress the efficiency of electron collection by the particle surface compared to that of the ions. The particle charge becomes less negative, i.e. decreases in the absolute magnitude compared to the case of an individual particle. In general, the relation between the particle charge and number density and the densities of electrons and ions in complex plasmas is governed by the quasineutrality condition and the competition between specific plasma production and loss mechanisms operating in a given situation. All this indicates that the consideration of an idealized Yukawa system with fixed particle charges and background plasma density can be in many cases insufficient to mimic the actual thermodynamics of real complex plasmas. How large modifications can be and whether they can be evaluated using conventional thermodynamic approaches require special careful investigation. We leave this for future work.

\begin{acknowledgments}
We thank Zoltan Donko for providing the values of sound velocities obtained using the QLCA model, which are listed in Table~\ref{Tab0}. We also thank Satoshi Hamaguchi for providing numerical data on wave dispersion relations in Yukawas fluids, partly reproduced in Figure~\ref{fig3}. This study was partially supported by the Russian Science Foundation, Project No. 14-12-01235.
\end{acknowledgments}

\appendix

\section{Polarization force on an individual particle} \label{Individual}

In the case of an individual particle, the only contribution to the energy is that from the sheath around the particle. The particle energy is therefore
$U=-T(\kappa\Gamma/2)=-Q^2/2\lambda_{\rm D}$. If the plasma is non-uniform such a particle will be acted upon by the force
\begin{equation}
F_{\rm pol}=-\nabla U=-\frac{Q^2\nabla \lambda_{\rm D}}{2\lambda_{\rm D}^2},
\end{equation}
which is known as the polarization force~\cite{Hamaguchi1,Hamaguchi2}. The polarization force is small in most practical situations occurring in complex (dusty) plasmas, but can affect considerably the dispersion of dust acoustic waves as has been pointed out in Ref.~\cite{KhrapakPRL2009}.

In the present consideration we treated the particle charge $Q$ constant. If this limitation is relaxed, then another contribution to the force appears formally
\begin{equation}
F_{Q}=\frac{Q\nabla Q}{\lambda_{\rm D}},
\end{equation}
which would push a particle in the region where its charge is higher. The question whether this is a real force or an artifact of this consideration and, in the first case, its significance for complex plasmas deserves separate detailed consideration.

\end{document}